\documentclass[RNAAS]{aastex62}
\usepackage{CJKutf8}

\graphicspath{{./}{figures/}}

\begin{document}

\title{A direct connection between the wake and the former host galaxy of a proposed runaway supermassive black hole}

\correspondingauthor{Pieter van Dokkum}
\email{pieter.vandokkum@yale.edu}

\author{Pieter van Dokkum}
\affiliation{Astronomy Department, Yale University, 52 Hillhouse Ave, New Haven, CT 06511}


\begin{abstract}

This Research Note presents VLT $B$-band imaging of a candidate runaway supermassive black hole
that was recently discovered in
HST/ACS F606W+F814W imaging. The ACS data
show an extremely thin, linear feature at $z=0.964$ that points toward a compact galaxy at
the same redshift.
There is a gap between the feature and the compact galaxy, which means that the proposed
causal connection between the two objects is not definitive. We show here that the
linear feature extends all the way to the compact
galaxy in the $B$-band, with no gap.  The $B$-band morphology is
difficult to reconcile with
models where the compact galaxy and the linear feature are independent objects, and in particular with
the proposal of \cite{s23} that the linear feature is an edge-on disk galaxy.

\end{abstract}

\gdef\blob{NGC\,1052-DF2}
\section{Introduction and motivation}
In \citet{vd23} (hereafter vD23) 
a remarkably thin, linear object  was described,
pointing to the center of a
compact galaxy (here dubbed GX) 
(see Fig.\ 1). Spectroscopy showed that the linear feature and 
GX are both at $z\approx 0.964$, and the feature was interpreted as the
wake of a runaway supermassive black hole that had escaped from GX some 40\,Myr before
the epoch of observation. Other explanations were considered, including the possibility that the
feature is an ultra-thin galaxy that is seen exactly edge-on. In a recent paper
\citet{s23} argue that the edge-on galaxy interpretation is not only possible but more
likely than a runaway black hole, largely based on the similarity between the velocity profile
along the feature and galaxy rotation curves.

These two scenarios imply different relationships between
the linear feature and GX.
In the runaway black hole model they are causally connected,
with the linear feature pointing to the center of GX because that is where the black hole came from.
By contrast, in the edge-on galaxy interpretation the alignment of the
feature with GX happened by chance; GX and the linear feature
are two galaxies in the same group but are otherwise independent. In fact, they have to be
some distance apart along the line of sight
to prevent heating of the disk \citep[see, e.g.,][]{selwood:14}.

\section{VLT/FORS2 $B$-band imaging}

As discussed in vD23, the then-available imaging and spectroscopy already suggested a
connection between the feature and GX: the feature appears to extend continuously from
GX to its head 62\,kpc away, without a gap,
in both the  [O\,{\sc iii}] line and the observed $u$ band
(see Fig.\ 6 in vD23).  However, the S/N ratio and spatial resolution of those observations are low,
and in this Research Note this possible connection is
investigated using high quality
$B$-band imaging.  The observations were obtained
with the FORS2 instrument on the VLT, in the context of program 386 in period 104 (PI:
M.\ Arnaboldi). They are centered on the dark matter-deficient galaxy NGC1052-DF2 \citep{vd18};
the linear feature is located near the edge of the field. We obtained the raw data
and calibration frames from the ESO Science Archive, and reduced them using
standard techniques. The three exposures with the best seeing were combined for a total
integration time of 1500\,s. The FWHM image quality in the combined image is $0\farcs 57$. 

The VLT $B$-band data are compared to the summed F606W+F814W HST/ACS image in Fig.\ 1. 
The morphology of the feature is different from that in the redder HST bands:
the pronounced gap between the feature and GX that is seen in the ACS bands is gone,
and the feature is continuous over its entire 62\,kpc extent. The flux in the
gap is significant at $>8\sigma$, and cannot be attributed to the resolution difference
between the VLT and ACS data.
The $B$ band samples $\lambda_{\rm rest}\approx 0.22\,\mu$m, and the VLT observations
confirm that the morphology of the feature in the rest-frame far-UV is distinct from
that in the rest-frame near-UV.
The rest-frame far-UV traces
both young stars and shocked gas, whereas the observed HST/ACS bands only
trace young stars (see vD23).
The VLT $B$-band image shows the shocked gas in the
ACS gap at higher spatial resolution and S/N ratio
than the CFHT $u$ band data of vD23.\footnote{There is
no evidence for a counter feature
on the other side of GX in the $B$-band image.  This could mean that the counter feature
that was found in vD23 is spurious or that the 
spectrum on that side is harder.}

\begin{figure}[htbp]
\begin{center}
\includegraphics[scale=0.4,angle=0]{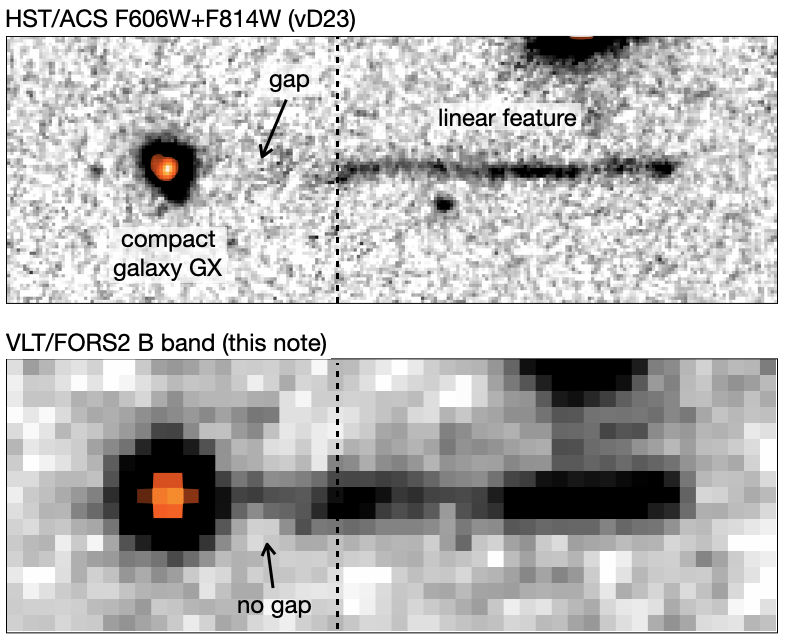}
\caption{Top: summed ACS F606W+F814W image
of the linear feature and the compact galaxy GX, both at $z=0.964$. The brightest emission
is in orange.
In \citet{vd23} the linear feature is interpreted as young stars
behind a runaway supermassive black hole that originated in
GX. There is a gap
between the feature and GX, and \citet{s23} explore the alternative
interpretation that they are two independent galaxies.
Bottom: VLT $B$ band image, sampling $\lambda_{\rm rest}\sim 0.22\,\mu$m. In the rest-frame
far-UV there is no gap and the feature extends all the way to GX. This is consistent with
what was seen in the observed $u$ band and the [O\,{\sc iii}] line in
vD23, and difficult to reconcile with the
\citet{s23} hypothesis. In the runaway
black hole model the emission in the gap is shocked gas.}
\end{center}
\end{figure}

\section{Implications for the edge-on galaxy hypothesis}

The VLT image in Fig.\ 1
shows that the linear feature and the
compact galaxy GX are connected in the rest-frame far-UV. The most straightforward
interpretation is that the alignment of the linear feature in the ACS data
with the center of GX is not a coincidence, and that they form a single system.
The far-UV morphology cannot be easily reconciled with the edge-on galaxy interpretation of the ACS morphology.

\section{Discussion}

Independent of morphological arguments, the emission line strengths
along the feature also strongly disfavor the
edge-on galaxy hypothesis. First, there is the extremely bright 
[O\,{\sc iii}] knot that is interpreted as the leading edge of the wake in vD23.
Its location at the tip of the feature is unexplained in the \citet{s23} 
model, as is its luminosity of
$L\approx 1.9\times 10^{41}$\,ergs\,s$^{-1}$. This corresponds to $5\times 10^7$\,L$_{\odot}$,
higher than the galaxy-integrated [O\,{\sc iii}] luminosities of virtually
all non-AGNs in the \citet{kewley:06} SDSS sample (see their Fig.\ 16).
Second, the [O\,{\sc iii}]/H$\beta$
line ratio reaches values of $\sim 10$ in two regions along the feature, including
the luminous knot at the head. Such high ratios
can only be explained with an active black hole or a strong shock with a velocity
$>500$\,km\,s$^{-1}$ \citep[see][and vD23]{allen:08}. 

The main argument in favor of the edge-on galaxy interpretation is that the velocities
along the (ACS) feature resemble a rotation curve, and the similarity of the velocity
profile to the rotation curve of IC\,5249 in Fig.\ 1 of \citet{s23} is indeed striking.
However, the velocity profile also shows a strong similarity to the
spatial displacement of the feature in the ACS data, as demonstrated in Fig.\ 10
of vD23. The excellent fit of the velocity profile to the ``wiggles'' in the ACS
data in the runaway black hole model (compare the red line to the data points in
Fig.\ 10 of vD23) would have to be coincidental in the edge-on galaxy model.
The correspondence between the color variation along the feature  and a simple
aging model (Fig.\ 9 of vD23) would also have to be a coincidence. \citet{s23}
suggest that the luminosity profile along the feature is inconsistent with linear aging, but
that is only true for a constant star formation rate;
in the black hole model,
the irregular luminosity profile
along the wake simply implies variation in the star formation rate.\footnote{As noted in vD23 there is some evidence for
episodic star formation; the typical separation between peaks is $\approx 4$\,kpc,
corresponding to $\approx 2$\,Myr.}

The interpretation of this object will likely remain a topic of debate for some time. The particular question
raised in \citet{s23} will likely be answered definitively in the Cycle 30 program
HST-GO-17301, which will provide very deep far-UV and white light imaging of the feature with UVIS on HST.




\acknowledgments
We thank M.\ Arnaboldi and collaborators for obtaining their excellent $B$-band image
of NGC1052-DF2. Based on data obtained from the ESO Science Archive Facility.

\end{document}